\documentclass[jkps,preprint,fleqn,showkeys]{revtex4}
\usepackage{graphicx}
\usepackage{amssymb}
\usepackage{amsmath}
\usepackage{bm}
\usepackage{threeparttable}
\usepackage{multirow}
\begin{document}
\setcounter{page}{0}
\title[]{Semi-empirical model to determine pre- and post-neutron fission product yields and neutron multiplicity}
\author{Jounghwa \surname{Lee}}
\affiliation{Nuclear Physics Application Research Division, Korea Atomic Energy Research Institute, Daejeon 34057, Korea}
\author{Young-Ouk \surname{Lee}}
\affiliation{Nuclear Physics Application Research Division, Korea Atomic Energy Research Institute, Daejeon 34057, Korea}
\author{Tae-Sun \surname{Park}}
\affiliation{Center for Exotic Nuclei Studies, Institute for Basic Science, Daejeon 34126, Korea}
\author{Peter \surname{Schillebeeckx}}
\affiliation{European Commission Joint Research Centre, Directorate G, Retieseweg 111, 2440 Geel, Belgium}
\author{Seung-Woo \surname{Hong}}
\email{swhong@skku.ac.kr}
\affiliation{Department of Physics, Sungkyunkwan University, Suwon 16419, Korea}

\date[]{Received 6 January 2021}

\begin{abstract}
Post-neutron emission fission product mass distributions
are calculated by using pre-neutron emission fission product yields (FPYs)
and neutron multiplicity.
A semi-empirical model is used
to calculate the pre-neutron FPY, first.
Then the neutron multiplicity for
each fission fragment mass is used 
to convert the pre-neutron FPY to
the post-neutron FPY.
In doing so, assumptions are made
for the probability
for a pre-emission fission fragment with a mass number $A^*$
to decay to a post-emission fragment with a mass number $A$.
The resulting post-neutron FPYs are compared
with the data available.
The systems where the experimental data of not only
the pre- and post-neutron FPY but also neutron multiplicity are available are
the thermal neutron-induced fission of
$^{233}$U, $^{235}$U and $^{239}$Pu.
Thus, we applied the model calculations to these systems
and compared the calculation results with those from the GEF
and the data from the ENDF and the EXFOR libraries.
Both the pre- and post-neutron fission product
mass distributions calculated by using 
the semi-empirical model and
the neutron multiplicity
reproduce the overall features of the experimental data.
\end{abstract}

\keywords{Fission product yield, Neutron multiplicity, Poisson distribution}

\maketitle

\section{INTRODUCTION}

Fission observables such as fission product yields (FPYs),
neutron multiplicity, and kinetic energy distributions 
are important quantities to understand the dynamics of fission.
In particular, the fission fragment mass distribution
is one of the most frequently measured observables.
Right after scission, fission fragments are highly neutron-rich
and de-excite through the emission of neutron and gamma rays.
The fission products before and after the emission of neutrons,
referred to as {\it pre-}neutron and {\it post-}neutron 
emission fission fragments, respectively, 
can be related through the emitted neutrons,
i.e., neutron multiplicity
if we consider only binary fission.
To have a comprehensive understanding of the fission process, 
one needs to describe consistently both pre- and post-neutron FPY data 
by taking into account the neutron multiplicity.
For post-neutron emission FPY (POST-FPY), many data are accumulated, 
and evaluated fission yield data are available 
for a wide range of fission systems 
in the evaluated libraries such as the ENDF \cite{Cha11}
and the JEFF \cite{Plo20},
but there are only a few experiments 
in which the pre-neutron emission FPY (PRE-FPY)
is measured (indirectly).
Theoretically, many studies of fission fragments
including microscopic approaches \cite{Gou05,Sca15,Reg16,Zha19}
and stochastic approaches \cite{Kar01,Asa04,Ran11,Ari13,Maz15}
were made to calculate the PRE-FPY through the evolution
of a compound nucleus
whereas the POST-FPY was rarely investigated, 
which is important in the applications such as
the calculation of decay heat or prediction of
fission fragment inventory.
Recently, there was a study \cite{Oku18} where the POST-FPY was calculated
from the PRE-FPY by using the Hauser-Feshbach model.
The Hauser-Feshbach model can treat 
the statistical decay of nuclei by particle emission,
but still relies on empirical models for initial conditions
such as spin and parity distribution and excitation energy distribution, 
which are essential for computation but not well known.
On the other hand, if data are available for the neutron multiplicities,
a data-driven model can be useful in calculating the POST-FPY
based on the multiplicity data.
For such a purpose, we attempt to calculate the POST-FPY
from the PRE-FPY by using neutron multiplicity data.

A simple semi-empirical model 
was proposed recently for the calculation of
the POST-FPY of uranium \cite{Lee18} and 
plutonium \cite{Lee20} isotopes
by taking into account the emission of prompt neutrons 
through the {\it average} number of neutrons emitted
for each {\it compound} nucleus, 
whereas the number of emitted neutrons 
depends on each individual fission fragment.
Therefore, in this work we attempt to calculate both 
PRE- and POST-FPYs with our semi-empirical fission model
by taking into account the number of emitted neutrons 
for each individual fission fragment mass.
We show that both the PRE- and  
POST-FPYs can be reproduced consistently 
by using the neutron multiplicity data.
For that, the number of neutrons emitted from 
{\it each} fission fragment mass is needed. 
We thus consider only the cases of 
thermal neutron-induced fission of
$^{233}$U, $^{235}$U and $^{239}$Pu, 
where the experimental data are available for
both PRE- and POST-FPYs
as well as the neutron multiplicity
for each fission fragment mass.
We first apply the semi-empirical FPY model to the
{\it pre}-neutron emission fission fragment mass distributions 
instead of the POST-FPY data as was done in Refs.~\cite{Lee18,Lee20}.
Then, the post-neutron fission fragment mass distributions 
can be calculated by taking into account the neutron multiplicity
for each fission fragment mass.
In doing so, the probability for pre-neutron emission fragments
with a mass number $A^*$ to decay to post-neutron emission fragments
with a mass number $A$ is needed.
We thus make assumptions for the probability in three different ways
based on the experimental neutron multiplicity data.
The POST-FPYs calculated by using the assumptions
are compared with the POST-FPYs from the experimental data 
\cite{Bid61,Far62,Thi76,Dii77,Mat77,Woh77,Weh80,Sch84,Qua88,Bai09,Gup17},
the ENDF/B-VII.1 \cite{Cha11}, and the GEF \cite{Sch16}.
It is shown that the PRE- and POST-FPYs calculated
by using the assumptions based on 
the experimental neutron multiplicity data can reproduce 
both PRE- and POST-FPY experimental data consistently.

\section{Model for pre-neutron FPY}

Let us begin with a brief description of the model \cite{Lee18}.
It is based on the work by
Bohr and Wheeler \cite{Boh39},
where the compound nucleus formed during the fission process
is considered as
a microcanonical ensemble.
The assumption that
the fission yield $Y(N,E^*)$
of fission fragments characterized by the
neutron number $N$
is determined by the level density at the fission barrier
leads us to \cite{Fon56,Sch08}
\begin{equation}
Y(N,E^*)=
Y_0\, \exp\left(2\sqrt{\tilde{a}\left(E^*-V(N)\right)}\right),
\label{YNAE}
\end{equation}
where $Y_0$ is an overall normalization factor,
$\tilde{a}$ is the level density parameter,
and $E^*=E_n+S(n)$ is the excitation energy of a compound nucleus,
deduced from the neutron separation energy $S(n)$
of the compound nucleus and the incident neutron energy $E_n$.
The fission barrier $V(N)$ in Eq.~(\ref{YNAE}) can be written as
\begin{equation}
V(N) = V_{mac}(N)+V_{sh}(N) \, \exp\left(-\gamma \varepsilon(N)\right)
\label{VN}
\end{equation}
with
\begin{equation}
\begin{split}
& V_{mac}(N) = C_{mac} \left(N - \frac{N_{CN}}{2}\right)^2 + V_0,
\\
& V_{sh}(N) = \\
&C_{in} \left[\exp\left(-\frac{(N-N_{in})^2}{\sigma_{in}^2}\right)+\exp\left(-\frac{(N-\bar{N}_{in})^2}{\sigma_{in}^2}\right)\right]
\\
& + C_{out} \left[\exp\left(-\frac{(N-N_{out})^2}{\sigma_{out}^2}\right)+\exp\left(-\frac{(N-\bar{N}_{out})^2}{\sigma_{out}^2}\right)\right],
\end{split}
\label{VV}
\end{equation}
where $V_0$ is a constant parameter,
$N_{CN}$ is the neutron number of the compound nucleus,
$N_{j}$ ($j=in,out$) determines the 
(inner and outer) peak positions of heavy fragments
taking care of shell effects,
and $\bar{N}_j$ ($\bar{N}_j\equiv N_{CN}-N_j$)
determines those of light fragments.
The factor $\exp\left(-\gamma \varepsilon(N)\right)$
in Eq.~(\ref{VN}) with the damping parameter $\gamma$
is for the quenching of the shell effects
at high excitation energies \cite{Ign75}
with
$\varepsilon(N) \equiv E^* - \left[ V_{mac}(N)+V_{sh}(N) \right]$.

In the previous works \cite{Lee18,Lee20}, 
the emission of prompt neutrons were taken into account 
in the fission barrier $V(N)$ by using 
the average number of emitted neutrons $\bar{\nu}$
known for each compound nucleus.
Therefore, the center of the fission yield distribution and
the light fragment peaks were shifted by $\bar{\nu}$
through $\tilde{N}_{CN}\equiv N_{CN}-\bar{\nu}$ and
$\tilde{N}_j\equiv \tilde{N}_{CN}-N_j (j=in,out)$, respectively.
In other words, in Refs.~\cite{Lee18,Lee20},
where Eq.~(\ref{VV}) was applied to describe the POST-FPY,
$\tilde{N}_{CN}$ and $\tilde{N}_j$ were used
in Eq.~(\ref{VV}) in place of $N_{CN}$ and $\bar{N}_j$.
Then, the model was applied to reproduce the POST-FPY data
by using the unchanged charge density postulate
\begin{equation}
\frac{N}{A}=\frac{N_{CN}-\bar{\nu}}{A_{CN}-\bar{\nu}},
\label{NA}
\end{equation}
which allows us to express our calculation results 
as a function of the fragment mass $A$
instead of the fragment neutron number $N$
through the average value $\bar{\nu}$ of
the neutron multiplicity.

It is generally thought that the prompt neutrons are emitted
from the nascent fission products.
Primary fission product yields
before emitting the prompt neutrons would have mirror 
symmetric shapes because the yields of heavy fragments and those of
the counter-part light fragments are expected to be the same
except for the rare cases of ternary fission. 
Thus, it is more appropriate to apply Eqs.~(\ref{YNAE})$\sim$(\ref{VV}), 
which would give us mirror symmetric yields,
to the {\it pre}-neutron emission FPY
rather than the POST-FPY.
Also, if the neutron multiplicity is available for
each fission fragment mass, 
as can be obtained from the experimental 
data \cite{Nis98a,Fra66,Nis98b,Vor10,Goo18,Adi20,Nis95,Tsu00},
we can treat the neutron multiplicity as a function of 
each fission fragment mass as $\nu=\nu(A^*)$.
By doing so, we may attempt to describe both the PRE- and
POST-FPYs in a consistent manner.
Because we apply  Eqs.~(\ref{YNAE})$\sim$(\ref{VV}) 
to the PRE-FPY, Eq.~(\ref{NA})
needs to be changed to
$N/A=N_{CN}/A_{CN}$ in converting the FPY
from a function of $N$ to that of $A$.
Unfortunately, there are only a few PRE-FPY data available, 
while quite a few POST-FPY data exist.
We also need the neutron multiplicity data $\nu(A^*)$ 
for each fragment mass.
The systems where both PRE- and POST-FPY data are available 
together with the neutron multiplicity data $\nu(A^*)$ are just
$^{233}$U(n$_{th}$,f), $^{235}$U(n$_{th}$,f) and $^{239}$Pu(n$_{th}$,f).
Thus, we only consider these three systems 
in applying the model.

\section{Model parameters to reproduce the pre-neutron emission FPY}

As seen in Eqs.~(\ref{YNAE})$\sim$(\ref{VV}), our model has ten parameters
in total:
$C_{mac}$, $C_{in}$, $C_{out}$, $V_0$,
$N_{in}$, $N_{out}$, $\sigma_{in}$, $\sigma_{out}$, $\gamma$ 
and $\tilde{a}$.
$C_{mac}$ determines the curvature of the macroscopic part of the fission 
barrier, $V_{mac}(N)$. 
We used Eqs.~(8) and (9) of Ref.~\cite{Ben98} for $C_{mac}$
that was fixed from the width of fission fragment mass distribution
at high excitation energies \cite{Itk88}.
Since the Gaussian functions of $V_{sh}(N)$ in Eq.~(\ref{VV}) take care of
modulation of the fission barrier due to neutron shell effects,
we fix the value of $N_{in}$ as 82 to take into account
the shell effect of
spherical nuclei and treat $N_{out}$ as an adjustable parameter.
Although the fission barrier heights of different isotopes may vary,
the same value of $V_0=5$ MeV may be used
for all uranium and plutonium isotopes as in Refs.~\cite{Lee18}
and \cite{Lee20}, respectively.
From the earlier analysis of level density parameters of nuclides,
the damping parameter $\gamma$ was found to be  
$(0.05\sim0.06$) MeV$^{-1}$ \cite{Ilj92}
which was consistent with the results ($\gamma=0.05$ MeV$^{-1}$)
of a microscopic shell model calculation \cite{Ign75}.
We use the value of $\gamma=0.06$ MeV$^{-1}$ by
following Ref.~\cite{Egi05}.

Therefore, only six parameters
($C_{in}$, $C_{out}$, $N_{out}$, $\sigma_{in}$, $\sigma_{out}$ and
$\tilde{a}$) are left and they are determined by fitting 
the {\it pre-}neutron emission FPY data.
As noted in Sect.II, in Refs.~\cite{Lee18,Lee20} the model was initially 
applied to the {\it post-}neutron FPY data
by using the average values of
neutron multiplicity ($\bar{\nu}$) for each compound nucleus. 
However, in this work we apply the model to the PRE-FPY data
and calculate the POST-FPY by using the experimental 
neutron multiplicity $\nu(A^*)$ available for each fission fragment mass
to be discussed in the next section.
Nuclear data libraries such as the ENDF and JEFF 
provide recommended values only for the POST-FPY.
The experimental data of the PRE-FPY
\cite{Nis95,Tsu00,Sch92,Sur72,Gel86,Bab97,Boh69,Dya69,Sim90,Wag96,Zey06,Aki71}
have been measured by using different methods,
most of which are based on the 2E method.
The 2E method requires the average number of neutrons
as a function of the fragment masses in the analysis.
The most accurate PRE-FPY experimental data available
are those in Ref.~\cite{Gel86}
taken by the 2E-1V method for $^{233}$U, $^{235}$U, and $^{239}$Pu,
and those in Ref.~\cite{Nis95}
taken by the 2E-2V method for $^{239}$Pu.
Therefore, the PRE-FPY data in Refs.~\cite{Nis95,Gel86}
are employed to find the model parameters.
The six adjustable parameters are determined
by minimizing $\langle \Delta Y^2\rangle$
defined by
\begin{equation}
\langle \Delta Y^2 \rangle
\equiv \frac{1}{i_{max}}\sum_{i=1}^{i_{max}} (Y_i-\bar{Y}_i)^2,
\end{equation}
where $i_{max}$ is the number of the data points,
$Y_i$ is the calculated yield, and $\bar{Y}_i$ is the
experimental yields from Refs.~\cite{Nis95,Gel86}.
The resulting ten parameters of the model are summarized 
in Table~\ref{para-tab}.
For comparison, the parameters obtained earlier by reproducing
the POST-FPY for uranium \cite{Lee18} and plutonium \cite{Lee20}
isotopes are also listed.

\begin{table*}
\centering
\caption{The model parameters 
listed in the middle column
are determined to reproduce the 
PRE-FPY of thermal neutron-induced fission of 
$^{233}$U, $^{235}$U and $^{239}$Pu, and are compared with
those obtained earlier \cite{Lee18,Lee20} 
to reproduce the POST-FPY.}
\begin{tabular}{c|ccccc|cc} 
\hline 
 & $^{233}$U & & $^{235}$U & & $^{239}$Pu & U (post) [Ref.~\cite{Lee18}] & Pu (post) [Ref.~\cite{Lee20}]\\
\hline
$N_{out}$ & \multicolumn{3}{c}{91.36} & & 93.02 & 90.50 & 91.08\\
$C_{in}\: $(MeV)  & \multicolumn{3}{c}{$-7.83$} & & $-8.43$ & $-9.31$ & $-10.9$ \\ 
$C_{out}\: $(MeV) & \multicolumn{3}{c}{$-15.4$} & & $-16.0$ & $-14.2$ & $-16.5$ \\
$\tilde{a}\: $(MeV$^{-1}$) & 72.5 & & 60.7 & & 48.6 & $11.6(241-A_{CN})$ & $ 2.50(256-A_{CN})$ \\
$\sigma_{in}$  & 7.28 & & 6.24 & & 6.99 & $0.662(248-A_{CN})$ & $0.51(254-A_{CN})$\\
$\sigma_{out}$ & 5.40 & & 5.97 & & 7.39 & 5.13 & 5.38\\ \hline
$V_0$ (MeV) & \multicolumn{7}{c}{5.0}  \\ 
$\gamma\: $(MeV$^{-1}$) & \multicolumn{7}{c}{0.06} \\ 
$N_{in}$ & \multicolumn{7}{c}{82} \\
$C_{mac}\: $(MeV) & \multicolumn{7}{c}{Eqs.~(8) and (9) of Ref.~\cite{Ben98}} \\ \hline
\end{tabular}
\label{para-tab}
\end{table*}

\begin{figure*}
\centering
\resizebox{0.9\textwidth}{!}{%
  \includegraphics{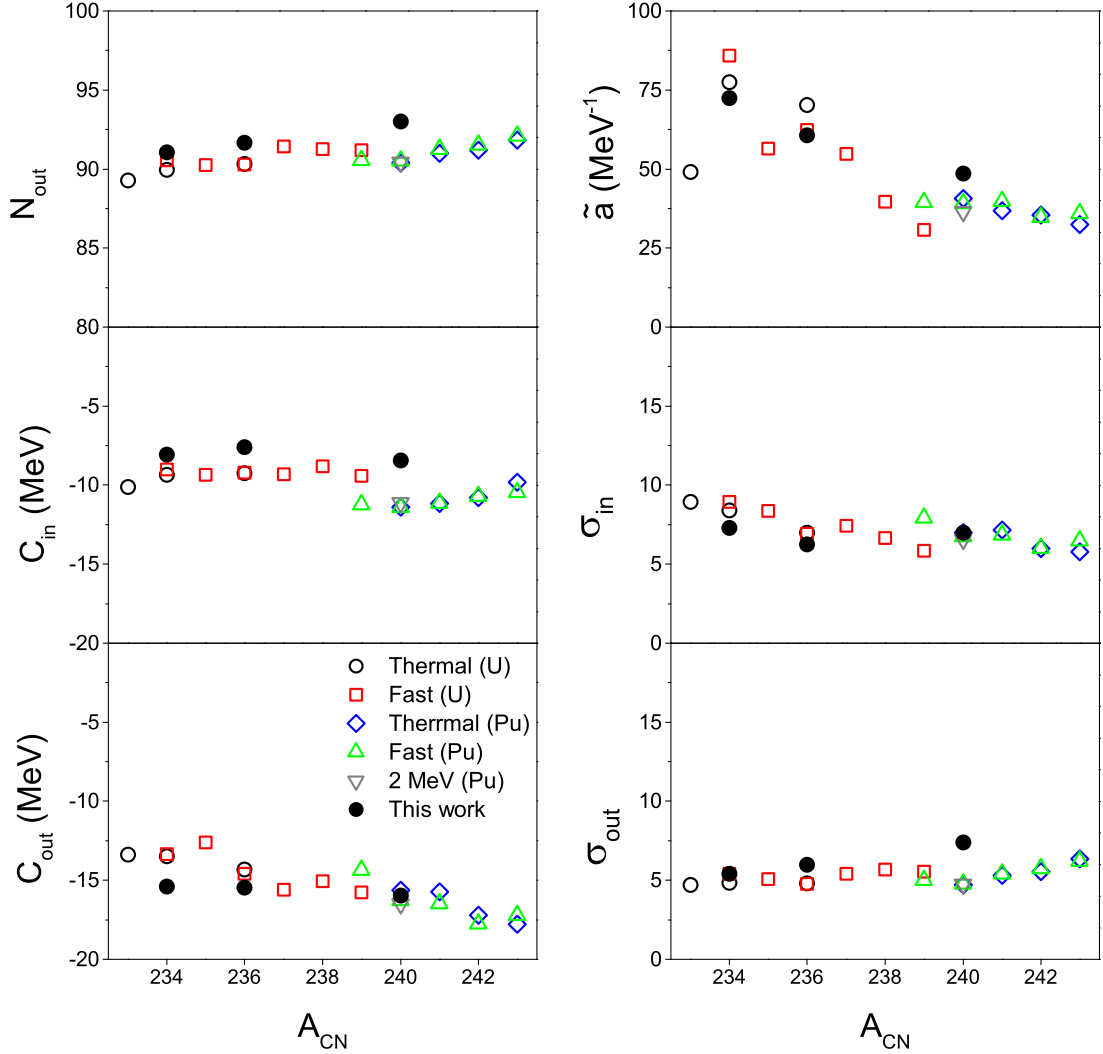}
}
\caption{The six parameters extracted to reproduce
the {\it pre}-neutron FPY of thermal neutron-induced fission of
$^{233}$U, $^{235}$U and $^{239}$Pu are compared with those of 
the previous works \cite{Lee18,Lee20} obtained to reproduce 
the {\it post}-neutron FPY data of uranium and plutonium isotopes.
The black solid circles represent the parameters obtained in this work
for uranium and plutonium isotopes.
The various open symbols represent the parameters extracted
in Refs.~\cite{Lee18,Lee20}.
The black empty circles and red empty squares denote the parameters
for reproducing the POST-FPY due to thermal and 500 keV
neutron-induced fission of uranium isotopes, respectively.
The blue rhombuses, green triangles and gray reversed
triangles represent, respectively, those for thermal, 500 keV and 2 MeV 
neutron-induced fission of plutonium isotopes.}
\label{para-fig}
\end{figure*}

\begin{figure*}
\centering
\resizebox{0.9\textwidth}{!}{%
  \includegraphics{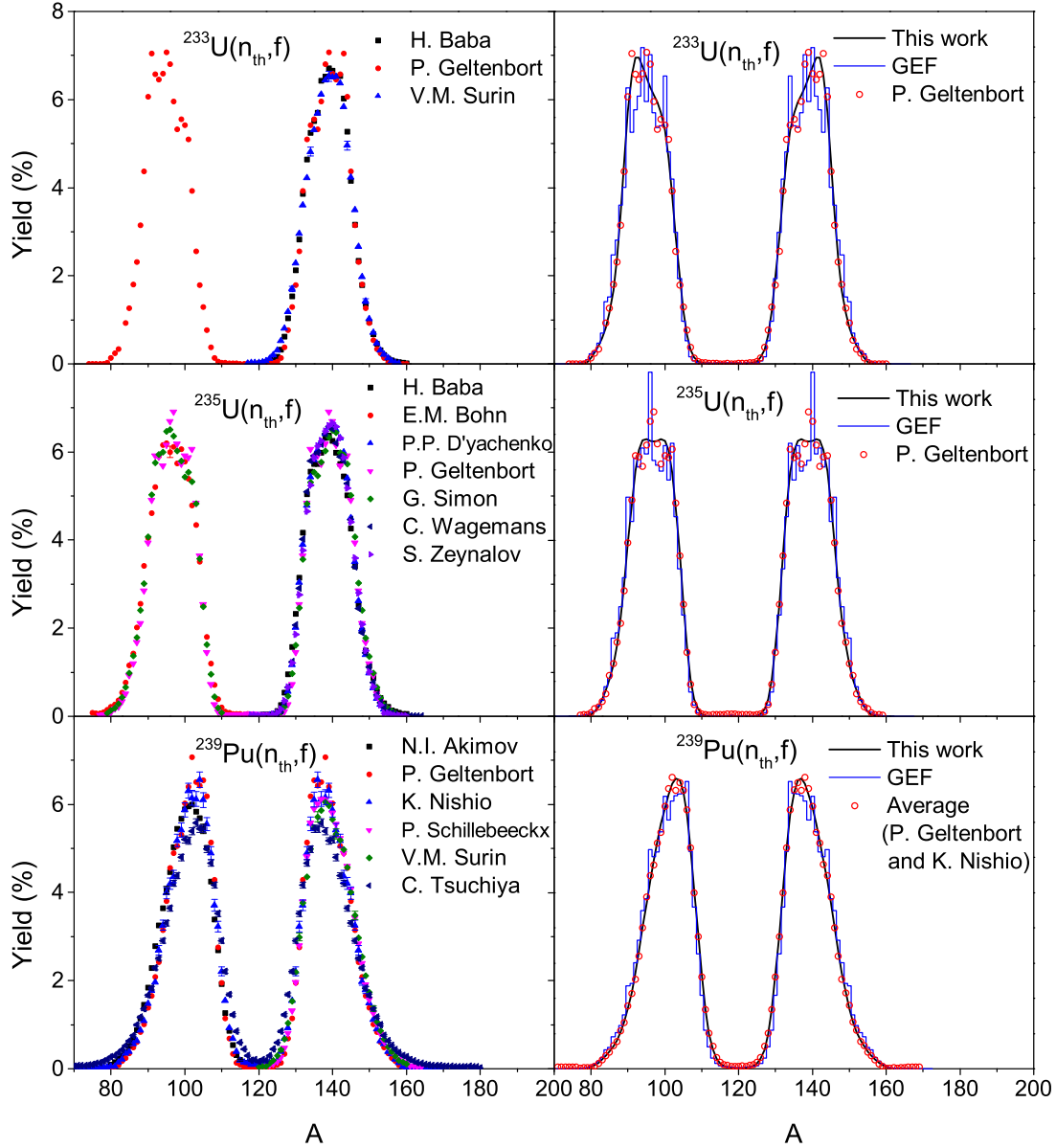}
}
\caption{Our calculation results for 
the PRE-FPY of
$^{233}$U, $^{235}$U and $^{239}$Pu by using the parameters in
Table~\ref{para-tab} are plotted on the right panels 
by the black full curves.
The experimental data \cite{Nis95,Gel86} are denoted
by the red circles on the right panels.
The PRE-FPY obtained by the GEF is plotted by
the blue histogram.
All the experimental PRE-FPY data 
available from the EXFOR
are plotted on the left panels for comparison.}
\label{pre-neutron}
\end{figure*}

In Fig.~\ref{para-fig}, the six parameters fixed 
by the least square fit of the PRE-FPY for 
$^{233}$U(n$_{th}$,f), $^{235}$U(n$_{th}$,f) and $^{239}$Pu(n$_{th}$,f)
are plotted together with those fixed in the previous works applied 
to the POST-FPY of uranium and plutonium isotopes. 
The model parameters obtained to reproduce 
the PRE-FPY data are in general similar to those \cite{Lee18,Lee20} 
obtained to reproduce the POST-FPY
except for $^{239}$Pu(n$_{th}$,f).
The parameters of the present work for 
$^{233}$U and $^{235}$U are quite
similar to those obtained for the POST-FPY of uranium isotopes.
The PRE-FPYs calculated with the parameters 
in Table~\ref{para-tab} are plotted by the black solid curves
in the right panels of Fig.~\ref{pre-neutron} 
together with those calculated by the GEF
which also provides the pre-neutron fission product mass distribution.
Because the GEF takes into account the even-odd effects 
in the proton and neutron number of fission fragments, 
there are spikes in the results of the GEF.
All the experimental data 
as well as the data \cite{Nis95,Gel86} used in the analyses are plotted
on the left panels.
The experimental data for $^{233}$U and $^{235}$U
of Ref.~\cite{Gel86}
are denoted by the red circles
in the upper right and middle right panels.
The average values of the experimental data for $^{239}$Pu
from Refs.~\cite{Nis95,Gel86} are
plotted by the red circles in the lower right panel.
The values of $\langle\Delta Y^2\rangle$ calculated from
the GEF and our model are listed in Table~\ref{pre-tab}.
Because our model parameters are fitted to reproduce 
the PRE-FPY data by minimizing $\langle\Delta Y^2\rangle$,
the values of $\langle\Delta Y^2\rangle$ from our model
are much smaller than those from the GEF.

\begin{table}
\centering
\caption{The values of $\langle \Delta Y^2\rangle$ for the PRE-FPY
calculated by the GEF and our semi-empirical fission model}
\begin{tabular}{|c|c|c|c|} \hline
\multirow{2}{*}{Fission} & \multirow{2}{5em}{No. of data points} & \multirow{2}{*}{GEF} & \multirow{2}{2.5em}{This work} \\
& & & \\
\hline \hline
$^{233}$U(n$_{th}$,f) & 87 & 0.205 & 0.055 \\ \hline
$^{235}$U(n$_{th}$,f) & 83 & 0.134 & 0.049 \\ \hline
$^{239}$Pu(n$_{th}$,f) & 95 & 0.073 & 0.006 \\ \hline
Average & & 0.137 & 0.037 \\ \hline
\end{tabular}
\label{pre-tab}
\end{table}

\section{From pre- to post-neutron emission FPY}

\begin{figure}
\centering
\resizebox{0.4\textwidth}{!}{%
  \includegraphics{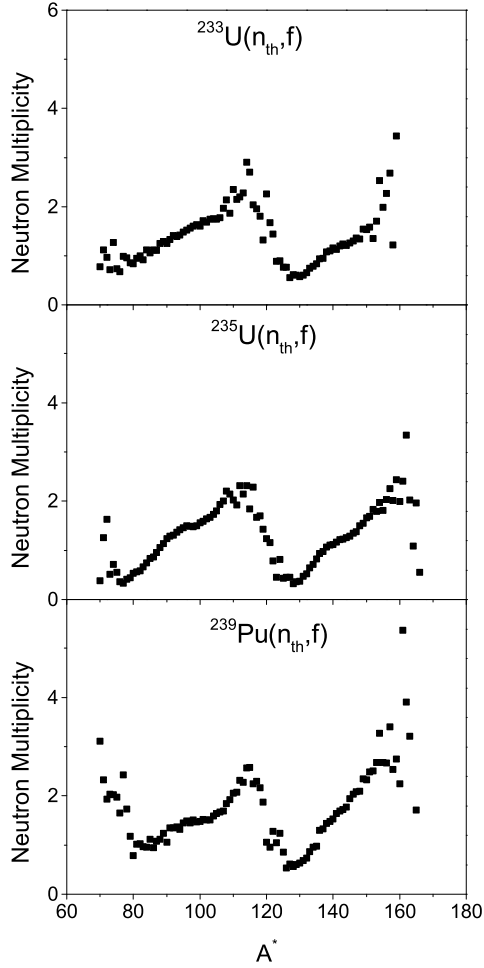}
}
\caption{The neutron multiplicities 
shown in the top panel for $^{233}$U
is from Ref.~\cite{Nis98a},
while those in the middle (for $^{235}$U)
and the bottom (for $^{239}$Pu) 
are the averaged values $\nu(A^*)^{avg}$.
(See the text for details.)
These neutron multiplicities are taken as $P_n(A^*)$ for
the 2-point approximation and averaged Poisson distribution.}
\label{mult_ave}
\end{figure}

Once the PRE-FPY is calculated, 
we can attempt to calculate the POST-FPY 
by using the neutron multiplicity data
available for $^{233}$U, $^{235}$U, and $^{239}$Pu.
The neutron multiplicities measured by different experimental groups
have discrepancies from each other.
Not all the measured neutron multiplicity data in the EXFOR \cite{Zer18}
have experimental uncertainties listed.
Thus, we took the averages of the experimental neutron multiplicities
for each fragment mass without considering the uncertainties.
The experimental values of neutron multiplicities with 
arbitrary units \cite{Kor71,Zam78}
or non-integer fragment masses \cite{Apa65}
are not included in the calculation.
Also, some experimental data sets have missing
neutron multiplicities for certain fragment masses.
Such data sets are not included in the calculation.
Then, we are left with only one set of 
experimental neutron multiplicity data
for $^{233}$U(n$_{th}$,f) \cite{Nis98a},
which is shown in the upper panel of Fig.~\ref{mult_ave}.
For $^{235}$U(n$_{th}$,f) and $^{239}$Pu(n$_{th}$,f), 
there are five \cite{Fra66,Nis98b,Vor10,Goo18,Adi20} 
and three \cite{Fra66,Nis95,Tsu00}
sets of data, respectively.
The experimental neutron multiplicities 
for $^{235}$U and $^{239}$Pu are averaged to get
$\nu(A^*)^{avg}$ and
are plotted in the middle and bottom panels of 
Fig.~\ref{mult_ave}, respectively.
Then, the POST-FPY $Y_{post}$ may be deduced 
from the PRE-FPY $Y_{pre}$ as follows.

Let $P_n(A^*)$ be the 
probability that a pre-neutron emission fragment
with a mass number $A^*$ decays to a post-neutron emission fragment
with a mass number $A=A^*-n$ by the emission of $n$ neutrons.
Then, $Y_{post}(A)$ can be written as
\begin{equation}
Y_{post}(A)=\sum_{n} Y_{pre}(A+n)P_n(A+n),
\label{Ypost}
\end{equation}
where $n=0,1,2,\dots$,
and the measured neutron multiplicity 
of a pre-fragment with a mass number $A^*$,
$\nu(A^*)$, would be given by
\begin{equation}
\nu(A^*)=\sum_{n} n P_n(A^*)
\label{nu}
\end{equation}
with the normalization condition
\begin{equation}
\sum_{n} P_n(A^*)=1.
\label{PAA}
\end{equation}
These relations are not sufficient to determine $P_n(A^*)$.
In fact, there was an experimental work to extract $P_n(A^*)$
for spontaneous fission of Cf and Fm isotopes \cite{Hof80},
where $P_n(A^*)$ seemed to have mostly Gaussian-shaped distributions
while for some cases with small values of neutron multiplicity
$P_n(A^*)$ seemed to have Poisson-shaped distributions.
(Note that for Cf and Fm isotopes, $\bar{\nu}$ is
relatively large as 3$\sim$4.)
$P_n(A^*)$ extracted in Ref.~\cite{Hof80} was not for each
fission fragment mass number $A^*$ but for small ranges of $A^*$.
Also, it was extracted from each fission event
with no distinction between light and heavy fragments.
On the other hand, $P_n(A^*)$
in Eqs.~(\ref{Ypost})$\sim$(\ref{PAA})
is for each fission fragment with a mass number $A^*$.
Thus, we resort to the following three assumptions for $P_n(A^*)$.

(1) 2-point distribution:	
$P_n(A^*)$ may be assumed to be
determined by two consecutive integer values of
$n_1$ and $n_1+1$ which are closest 
to the experimental neutron multiplicity values.
In other words, we assume that
\begin{eqnarray}
P_{n}^{2P}(A^*) = \left\{
\begin{array}{ll}
1 - r_1, & \text{for } n= n_1, \\
r_1, & \text{for } n= n_1 + 1,\\ 
0, & \text{for other } n,
\end{array}
\right.
\end{eqnarray}
where $n_1$ and $r_1$ are, respectively, the integer and
the fractional part of
the averaged neutron multiplicity, $\nu(A^*)^{avg}$
shown in Fig.~\ref{mult_ave}.
That is, $n_1 \equiv \left\lfloor \nu(A^*)^{avg} \right\rfloor$
and $r_1 \equiv \nu(A^*)^{avg} - \left\lfloor \nu(A^*)^{avg} \right\rfloor$,
where $\left\lfloor x \right\rfloor$
denotes 
the greatest integer less than or equal to 
a real number $x$.
We refer to this $P_{n}^{2P}(A^*)$ as the 2-point distribution.

(2) Averaged Poisson distribution:
We may assume that $P_n(A^*)$ follows the Poisson distribution
\begin{equation}
P_{n}^{PS}(\lambda)=\frac{\lambda^n e^{-\lambda}}{n!}
\label{Poisson}
\end{equation}
with the parameter $\lambda$ being the {\it average} value of $\nu(A^*)$.
We take the average experimental neutron multiplicities
shown in Fig.~\ref{mult_ave} for the value of $\lambda$.
The values of $\nu(A^*)^{avg}$ for
almost all fission fragments are less than 4,
and thus we may assume that $n$,
the number of emitted neutrons,
can run up to 10
for practical calculations
so that the cumulative Poisson distribution is larger than 0.99.

As an alternative to $P_{n}^{PS}(n;\lambda)$,
we also considered the binomial distribution
\begin{equation}
P_{n}^{BI}(n_{max},p)=\frac{n_{max}!}{n!(n_{max}-n)!} p^x (1-p)^{n_{max}-n},
\label{Binomial}
\end{equation}
where $n(=0, 1, 2, \dotsb, n_{max})$ corresponds
to the number of emitted neutrons
with another parameter $n_{max}$ and $p=\nu(A^*)^{avg}/n_{max}$.
However, we find that	
$Y_{post}$ calculated by using the binomial distribution 
$P_{n}^{BI}$ is practically the same as that from $P_{n}^{PS}$.
Therefore, the results by using the binomial distribution
are not shown here.

\begin{figure}
\centering
\resizebox{0.4\textwidth}{!}{%
  \includegraphics{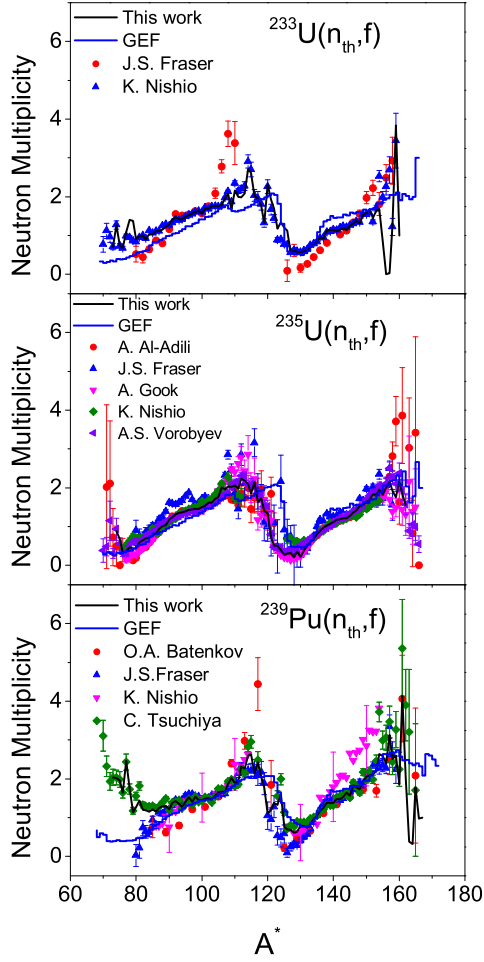}
}
\caption{The neutron multiplicities for
$^{233}$U(n$_{th}$,f), $^{235}$U(n$_{th}$,f) 
and $^{239}$Pu(n$_{th}$,f) obtained
by minimizing $\chi^2_{S}$ in Eq.~(\ref{chi})
with the optimized Poisson distribution for $P_n(A)$
are plotted by the black solid curves
in comparison with the experimental data
used in the calculations.
For comparison, the neutron multiplicities
calculated by the GEF
are presented by the blue step lines.}
\label{mult_po}
\end{figure}

(3) Optimized Poisson distribution:
Another way of determining $P_n(A^*)$ is 
to use the Poisson distribution of Eq.~(\ref{Poisson}),
but instead of using $\lambda=\nu(A^*)^{avg}$ 
we look for the values of $\lambda$ 
for each $A^*$ to minimize 
the value $\chi^2_{S}$ defined by
\begin{equation}
\begin{split}
\chi^2_{S} & =\chi^2_{Y}+\chi^2_{\nu} \\
& = \frac{1}{i_{max}} \sum_{i=1}^{i_{max}}\left(\frac{Y_i-Y^{exp}_i}{\Delta Y^{exp}_i}\right)^2+\frac{1}{j_{max}} \sum_{j=1}^{j_{max}}\left(\frac{\nu_{j}-\nu^{exp}_{j}}{\Delta \nu^{exp}_{j}}\right)^2.
\label{chi}
\end{split}
\end{equation}
In Eq.~(\ref{chi}), $i_{max}$ and $j_{max}$
are the number of yield data 
and neutron multiplicity data, respectively,
and $Y_i$ and $\nu_{j}$ are
the fission yield and neutron multiplicity
calculated by Eqs.~(\ref{Ypost}) and (\ref{nu}), respectively.
$Y^{exp}_i$ and $\Delta Y^{exp}_i$ 
are the fission yield data and the uncertainties from the ENDF/B-VII.1,
while $\nu^{exp}_{j}$ and $\Delta \nu^{exp}_{j}$ are 
the experimental neutron multiplicity data 
and their uncertainties shown in Fig.~\ref{mult_po}.
In this third method, $\lambda$ in Eq.~(\ref{Poisson}) is treated
as a free parameter determined by minimizing $\chi^2_{S}$
for each fission fragment mass number $A^*$.
The neutron multiplicities $\lambda$
determined in this way
are plotted in Fig.~\ref{mult_po} by the black solid lines,
which agree in general with the experimental data.
We denote such neutron multiplicity $\lambda$ by
$\nu(A^*)^{opt}$ to distinguish them from $\nu(A^*)^{avg}$.

\begin{figure*}
\centering
\resizebox{0.9\textwidth}{!}{%
  \includegraphics{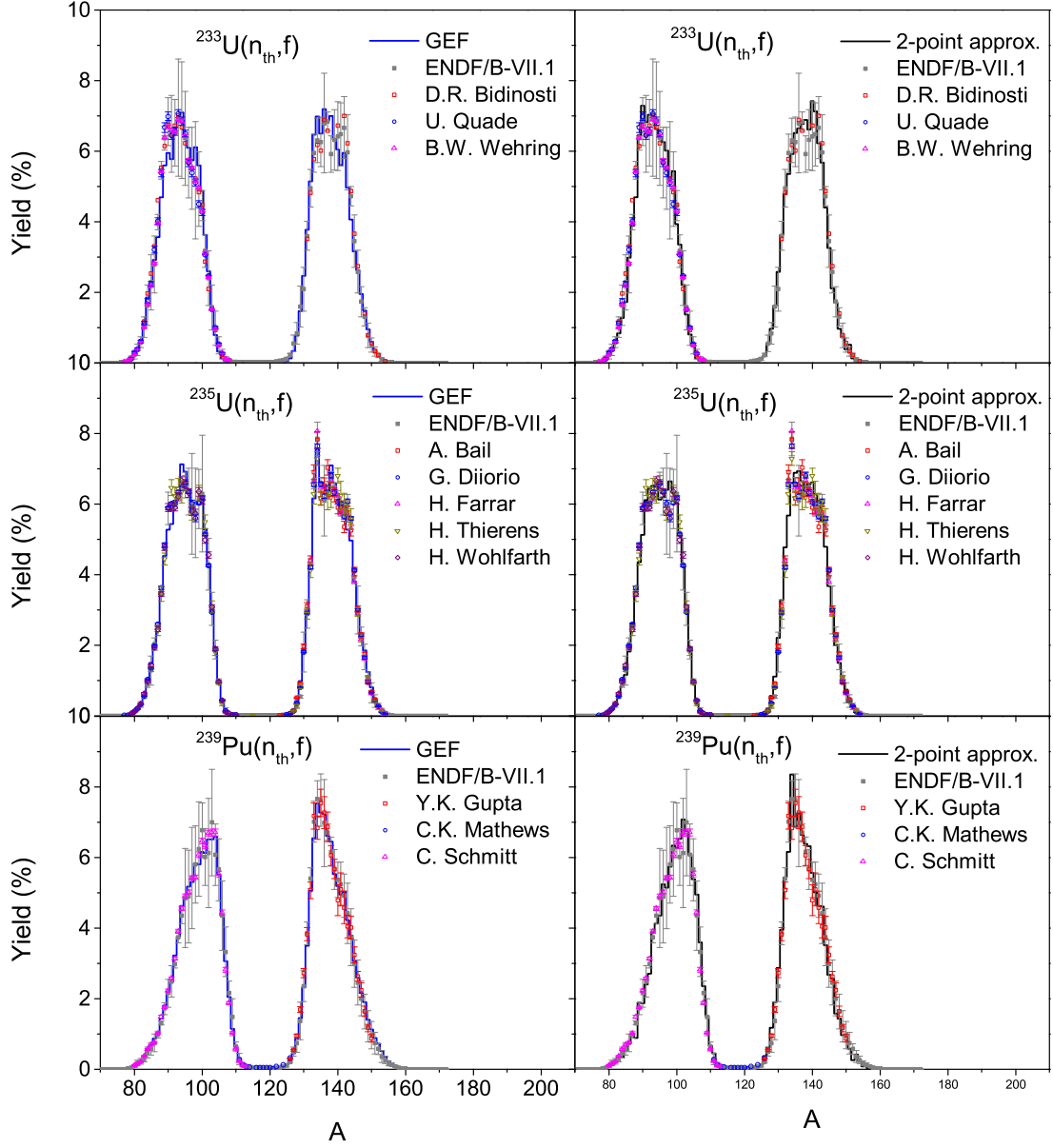}
}
\caption{The POST-FPY for 
the thermal neutron-induced fission
of $^{233}$U, $^{235}$U and $^{239}$Pu are plotted.
$Y_{post}(A)$ calculated by using the 2-point approximation (the black step lines)
is compared with the GEF (the blue step lines) and the
ENDF/B-VII.1 data (the gray dots with error bars).
Experimental data
\cite{Bid61,Far62,Thi76,Dii77,Mat77,Woh77,Weh80,Sch84,Qua88,Bai09,Gup17}
from the EXFOR library are represented by different symbols.}
\label{post-neutron_2p}
\end{figure*}

\begin{figure*}
\centering
\resizebox{0.9\textwidth}{!}{%
  \includegraphics{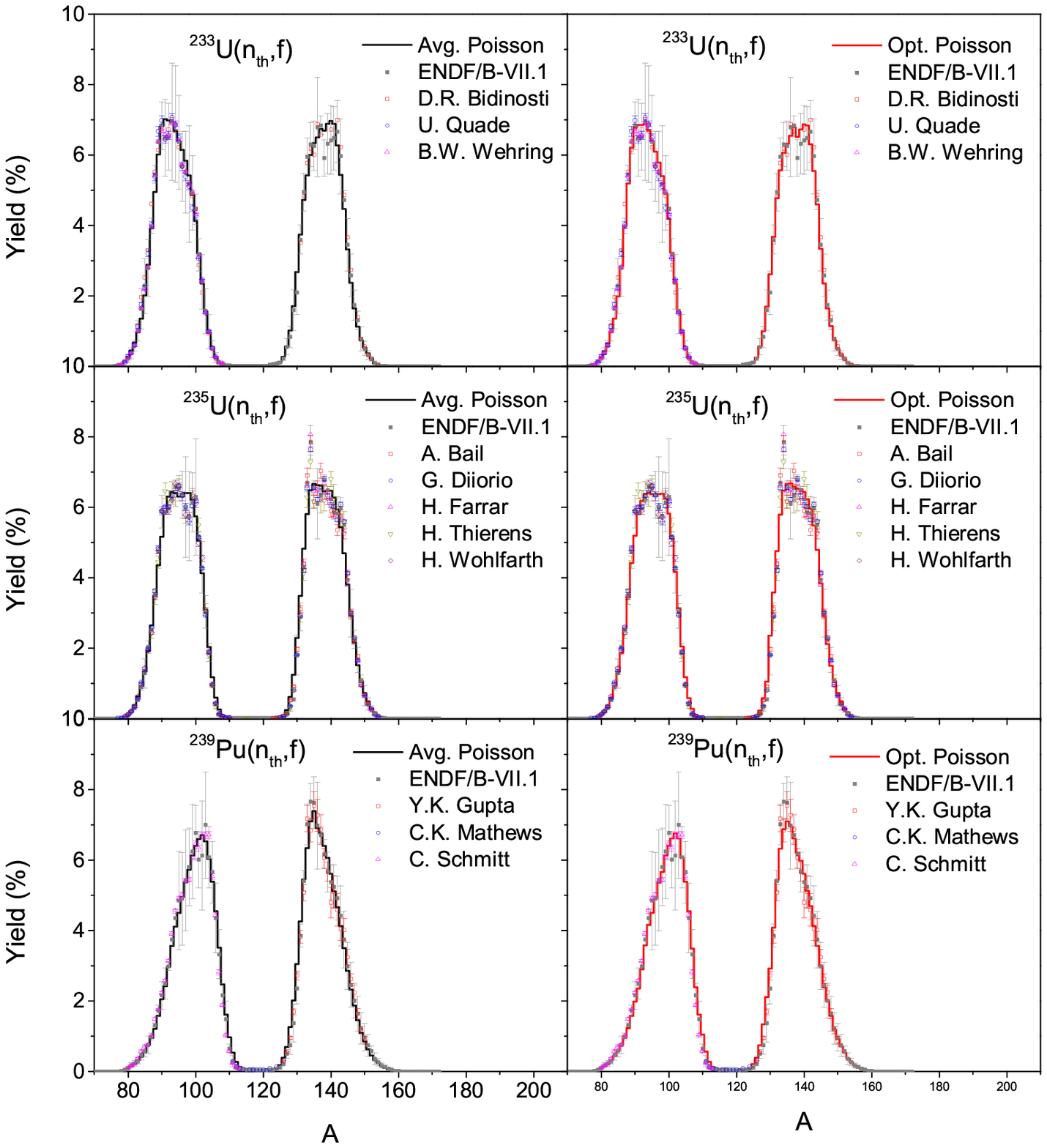}
}
\caption{The POST-FPY for 
the thermal neutron-induced fission
of $^{233}$U, $^{235}$U and $^{239}$Pu are plotted.
$Y_{post}(A)$ calculated by using 
the averaged Poisson distribution (the black step lines) are
plotted on the left panels in comparison with 
the ENDF/B-VII.1 data (the gray dots with error bars).
$Y_{post}$ calculated by using the optimized Poisson distribution
are plotted by the red step lines on the right panels.}
\label{post-neutron_po}
\end{figure*}

$Y_{post}(A)$s calculated in these three methods for $P_n(A^*)$
are compared with the experimental yield data 
and the ENDF evaluated data in
Figs.~\ref{post-neutron_2p} and \ref{post-neutron_po}.
$Y_{post}(A)$ calculated for $^{233}$U, $^{235}$U and
$^{239}$Pu by using the 2-point distribution
is presented in the right panels of Fig.~\ref{post-neutron_2p} 
by the black step lines 
together with the ENDF/B-VII.1 \cite{Cha11} data
plotted by the gray dots with error bars.
For comparison, the results 
from the GEF 2021/1.1 model \cite{Sch16}
are plotted on the left panels by the blue solid step lines.
The experimental data 
\cite{Bid61,Far62,Thi76,Dii77,Mat77,Woh77,Weh80,Sch84,Qua88,Bai09,Gup17}
from the EXFOR library
are denoted by different symbols
as indicated in Fig.~\ref{post-neutron_2p}.
$Y_{post}(A)$ calculated by using the averaged Poisson distribution
and the optimized Poisson distribution
are presented by the solid step lines
in the left and right panels of Fig.~\ref{post-neutron_po}, respectively.
Among the experimental data available in the EXFOR library,
those data taken by using 
radiochemical separations \cite{Bid61,Mat77},
mass spectrometers \cite{Far62,Dii77,Woh77,Weh80,Sch84,Qua88,Bai09,Gup17},
and catcher foil technique \cite{Thi76}
are known to be more accurate than those measured by the 2E method.
Thus, only the data from
Ref.~\cite{Bid61,Far62,Thi76,Dii77,Mat77,Woh77,Weh80,Sch84,Qua88,Bai09,Gup17}
are shown in Figs.~\ref{post-neutron_2p} and \ref{post-neutron_po}.

Due to the mirror symmetric characteristics 
of smooth functions in
Eqs.~(\ref{YNAE})$\sim$(\ref{VV}),
the detailed fluctuating structures in the POST-FPY
were not reproduced in Refs.~\cite{Lee18,Lee20}.
On the other hand, by using $Y_{pre}(A^*)$ calculated by
Eqs.~(\ref{YNAE})$\sim$(\ref{VV})
and taking into account
the neutron multiplicities, not only the overall shapes 
of the POST-FPY but also some detailed
structures near the peaks seem to be reproduced
better than in Refs.~\cite{Lee18,Lee20}.
The 2-point approximation seems very crude,
but $Y_{post}(A)$ calculated by the 2-point approximation
appears to reproduce the experimental POST-FPY data
as seen in Fig.~\ref{post-neutron_2p}.
The spike at $A=134$ of $^{239}$Pu(n$_{th}$,f)
seems reproduced by the 2-point approximation
as seen in the lower right panel of Fig.~\ref{post-neutron_2p}.
On the other hand, the Poisson distributions
with smooth mass distributions
result in $Y_{post}(A)$ with little fluctuating
structures as seen in Fig.~\ref{post-neutron_po}.

\begin{figure}
\centering
\resizebox{0.4\textwidth}{!}{%
  \includegraphics{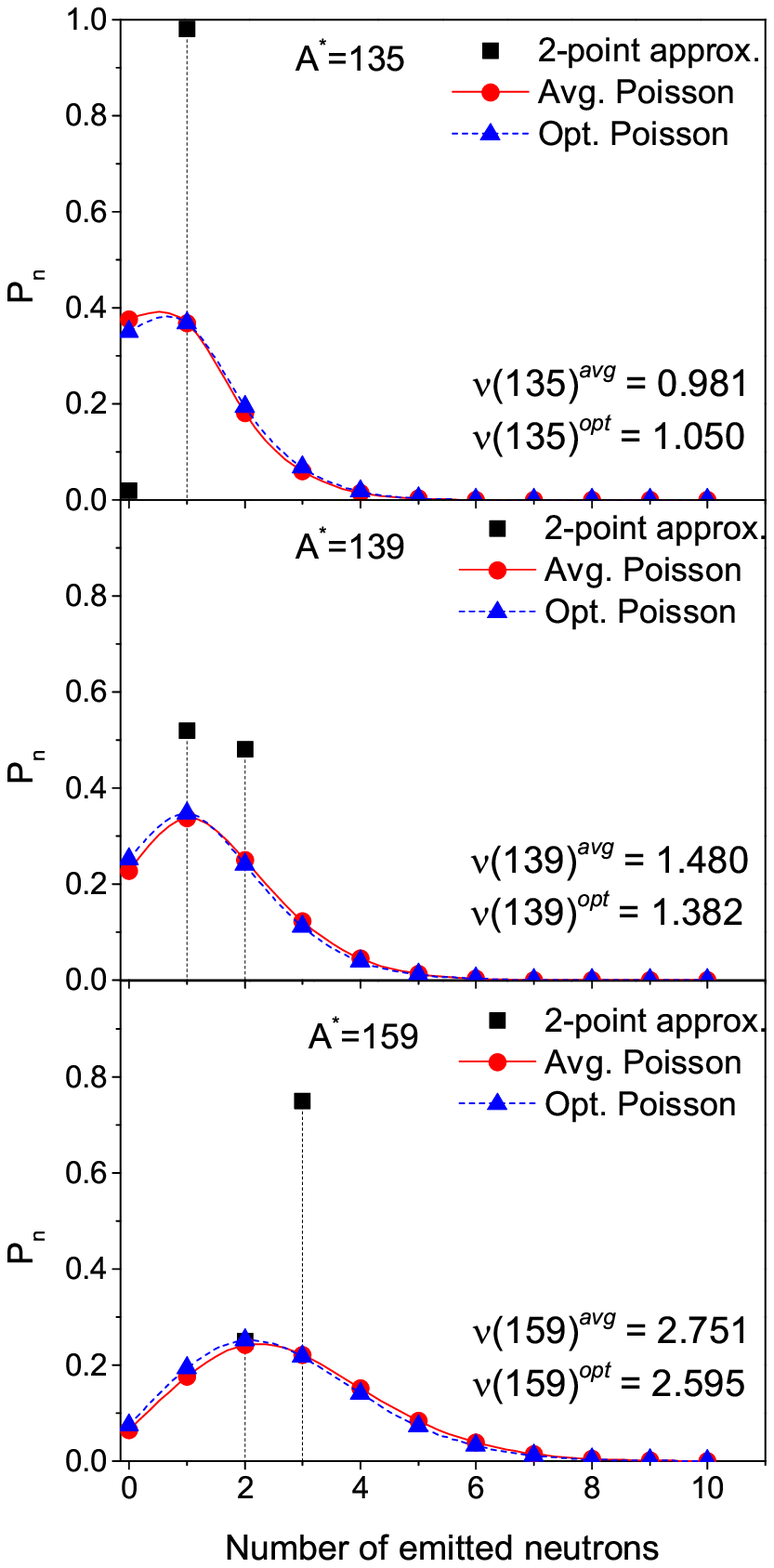}
}
\caption{$P_n(A^*)$ distributions are plotted for
the 2-point distribution, averaged Poisson and
optimized Poisson distribution for the cases of
$^{239}$Pu(n$_{th}$,f) with $A^*=135$ (upper panel), 
$A^*=139$ (middle panel) and $A^*=159$ (lower panel)
by the black squares, the red circles,
and the blue triangles, respectively.
$\nu(A^*)^{avg}$ and $\nu(A^*)^{opt}$ refer to
the average value of the experimental 
neutron multiplicity data shown in Fig.~\ref{mult_ave}
and the neutron multiplicity obtained by minimizing
the value of $\chi^2_s$ of Eq.~(\ref{chi}), respectively.}
\label{Pn_dist}
\end{figure}

In Fig.~\ref{Pn_dist}, $P_n(A^*)$ from the 
2-point distribution, averaged Poisson and
optimized Poisson distributions for $^{239}$Pu(n$_{th}$,f)
are plotted for $A^*=135, 139$ and 159
by the black squares, the red circles,
and the blue triangles, respectively.
These choices are made because they correspond to the cases
of $\nu(A^*)\approx 1$, 1.5 and 3.
When the value of $\nu(A^*=135)^{avg}=0.981$,
very close to 1,
$P_n$ is concentrated at $n=1$ in the 2-point distribution
as can be seen in the upper panel of Fig.~\ref{Pn_dist}.
As a result, most of the fission fragments with $A^*=135$ 
contribute to the generation of
the spike at $A=134$ of the POST-FPY of $^{239}$Pu
in the 2-point distribution
by emitting one neutron.
(See the bottom right panel of Fig.~\ref{post-neutron_2p}.)
On the other hand, the value of $P_n(A^*=135)$ from the averaged 
and optimized Poisson distributions
do not exceed 0.4
and have relatively broad distributions
as shown in the top panel of Fig.~\ref{Pn_dist},
which cannot generate sharp fluctuation structures in the POST-FPY.
As the value of $\nu(A^*)^{avg}$ increases, 
$P_n$s of both Poisson distributions
become broader and the peak positions are shifted to the right
as seen in the middle and lower panels of Fig.~\ref{Pn_dist}.
When $\nu(A^*=139)^{avg}=1.48$, close to 1.5,
$P_n$ is evenly distributed in the 2-point approximation
as seen in the middle panel of Fig.~\ref{Pn_dist}.
When $\nu(A^*=159)^{avg}=2.75$, close to 3,
the Poisson distributions are further shifted to the right
and become similar to a Gaussian distribution, as it should.

We remark that the spike at $A=134$
in the ENDF data for $^{235}$U is not reproduced
by the 2-point approximation as seen in the middle right panel
of Fig.~\ref{post-neutron_2p}
though a similar spike at $A=134$ is reproduced for $^{239}$Pu.
This is due to the use of the averaged value
of neutron multiplicity $\nu(A^*)^{avg}$.
If we use the experimental neutron multiplicity data 
of Fraser \cite{Fra66},
where $\nu(135)$ is 0.997, close to 1, 
then the spike at $A=134$ is again well reproduced.
Our results of the POST-FPY shown 
in Figs.~\ref{post-neutron_2p} and \ref{post-neutron_po}
seem to suggest that the 2-point approximation, 
being a naive, simple model,
seems useful in reproducing the spike particularly
when $\nu(A^*)$ is close to 1.
As $\nu(A^*)$ becomes larger, $P_n(A^*)$ seems to follow close to
a Poisson distribution as Fig.~\ref{Pn_dist} shows.
If $\nu(A^*)$ becomes even larger,
$P_n(A^*)$ seems to follow a Gaussian distribution, 
as Ref.~\cite{Hof80} shows.

To see quantitatively the degree of agreement between
the calculation results and the ENDF data,
we list in Table~\ref{post-tab} the values of $\chi^2_{Y}$
of Eq.~(\ref{chi}) and those of $\langle\Delta Y^2\rangle$
calculated by using the ENDF data
as the experimental POST-FPY.
Though our model parameters were fitted to reproduce 
the PRE-FPY data rather than the post-neutron data,
the values of $\langle\Delta Y^2\rangle$ calculated
for the POST-FPY from our model are similar to 
those from the GEF.
On the other hand, the values of $\chi^2_{Y}$ from our model
are somewhat larger than those from the GEF.
Note that the GEF is a sophisticated model
with about 50 parameters
while our model has essentially six adjustable parameters.
In the case of $^{235}$U, 
the uncertainties in the FPY data are small for all the fragments 
because of a large amount of accumulated experimental data.
Therefore, the values of $\chi^2_{Y}$ for $^{235}$U are larger than 
those for $^{233}$U and $^{239}$Pu in both our model and the GEF.
Note that the values of $\chi^2_{Y}$ 
from the optimized Poisson distribution are a little smaller
than those from the averaged Poisson distribution
while the two Poisson distributions look 
quite similar to each other as seen in Fig.~\ref{Pn_dist}.
Out of nearly 90 data points of $A^*$,
there are just about 5 data points for each isotope 
where the two Poisson distributions do not agree well,
though they are not shown here.
Those few points are all in the valley or in the wing regions,
where the evaluated uncertainties of yields are very small.
Due to the small uncertainties of yields,
the values of $\chi^2_{Y}$ from 
the two Poisson distributions are different,
though $P_n(A^*)$s look very similar
for most values of $A^*$.

At present, there are not many cases where PRE- and POST-FPY 
and neutron multiplicity data are available simultaneously.
There is no recommended value like the ENDF data for the PRE-FPY.
The neutron multiplicity values from different experimental groups
have discrepancies.
Thus, more experimental FPY and neutron multiplicity data are needed
to describe the PRE- and POST-FPY in a consistent way.

\begin{table*}
\centering
\caption{$\langle \Delta Y^2\rangle$ and $\chi^2_{Y}$ values calculated by the GEF and our model for the POST-FPY}
\begin{threeparttable}
\begin{tabular}{|c|c|r|r|r|r|r|r|r|r|} \hline
 & No. of & \multicolumn{4}{|c|}{$\langle\Delta Y^2\rangle$} & \multicolumn{4}{|c|}{$\chi^2_{Y}$} \\
\cline{3-10}
Fission & data & GEF & 2-point & \multicolumn{1}{|c|}{Avg.} & \multicolumn{1}{|c|}{Opt.} & GEF & 2-point & \multicolumn{1}{|c|}{Avg.} & \multicolumn{1}{|c|}{Opt.} \\ 
 & points & & approx. & \multicolumn{1}{|c|}{Poisson} & \multicolumn{1}{|c|}{Poisson} & & approx. & \multicolumn{1}{|c|}{Poisson} & \multicolumn{1}{|c|}{Poisson} \\
\hline \hline
$^{233}$U(n$_{th}$,f) & 85 & 0.1217 & 0.0848 & 0.0532 & 0.0337 & 1.17 & 1.92 & 1.40 & 1.02 \\ \hline
$^{235}$U(n$_{th}$,f) & 86 & 0.0361 & 0.0970 & 0.0911 & 0.0831 & 4.09 & 6.68 & 7.63 & 5.52 \\ \hline
$^{239}$Pu(n$_{th}$,f) & 88 & 0.0335 & 0.0786 & 0.0534 & 0.0562 & 0.40 & 2.08 & 1.13 & 1.24 \\ \hline
Average & & 0.0638 & 0.0868 & 0.0659 & 0.0576 & 1.89 & 3.56 & 3.39 & 2.60 \\
\hline
\end{tabular}
\end{threeparttable}
\label{post-tab}
\end{table*}

\section{Summary}

We applied a semi-empirical model to compute first the PRE-FPY
of thermal neutron-induced fission of $^{233}$U, $^{235}$U and $^{239}$Pu,
and then calculated the POST-FPY by 
using the neutron multiplicity data.
In our previous works \cite{Lee18,Lee20}, the semi-empirical model was
applied to the POST-FPY taken from the ``ENDF/B-VII.1".
In this work, however, the model parameters were fixed
to reproduce the experimental PRE-FPY data.
The six model parameters obtained to reproduce the PRE-FPY
were in general close to those obtained to reproduce
the POST-FPY except for $^{239}$Pu.
The POST-FPY are then calculated by using 
the model parameters fixed to the primary FPY 
and the neutron multiplicity data with three different assumptions
about the probability of the number of emitted neutrons
for each fission fragment mass number $A^*$:
2-point approximation, averaged Poisson distribution and
optimized Poisson distribution.
The resulting POST-FPYs obtained 
from the three different assumptions
reproduce the overall shapes
of the POST-FPY from the ENDF.
Although the 2-point distribution
is a very simple and naive assumption,
it shows a possibility to describe the fluctuating structures
of the POST-FPY.
In particular, when $\nu(A^*)$ is less than $\sim 1.5$,
$P_n(A^*)$ may be well approximated to the 2-point distribution.
As $\nu(A^*)$ increases,
$P_n(A^*)$ seems to follow the Poisson distribution.
If $\nu(A^*)$ further increases,
$P_n(A^*)$ seems to follow the Gaussian distribution.
Though the model parameters were determined 
to reproduce the pre-neutron data,
the values of $\langle\Delta Y^2\rangle$ and $\chi^2_{Y}$,
indicating the degree of agreement
between the calculated POST-FPY and the ENDF data,
are comparable to those of the GEF.
If more accurate neutron multiplicity data are available,
we can test the model better.

\begin{acknowledgments}
This work was supported in part by the National Research Foundation  (NRF) 
of Korea funded by the Ministry of Science and ICT (NRF-2020R1A2C1102384, NRF-2018M7A1A1072274). 
TSP acknowledges the support from the IBS grant funded by the Korean government (No. IBS-R031-D1).
\end{acknowledgments}

\end{document}